\documentclass[aps,prl,twocolumn,floats,amsmath,amssymb,showpacs,floatfix]
{revtex4}
\newcommand\be{\begin{equation}}
\newcommand\ee{\end{equation}}
\newcommand\bea{\begin{eqnarray}}
\newcommand\eea{\end{eqnarray}}
\newcommand\ket[1]{|#1\rangle}
\newcommand\bra[1]{\langle #1|}

\newcommand{\bk}[2]{{\langle #1|}{#2\rangle}}
\usepackage{amssymb,amsmath,palatino}
\usepackage{dcolumn}

\begin{document}

\title{Mutually Unbiased Bases for Continuous Variables}
\author{Stefan Weigert$^{1}$ and Michael Wilkinson$^{2}$}
\affiliation{
$^{1}$ Department of Mathematics, University of York, Heslington, York YO10 5DD, England\\$^{2}$Department of Mathematics and Statistics, The Open
University, Walton Hall,
Milton Keynes, MK7 6AA, England\\}

\begin{abstract}
The concept of mutually unbiased bases is studied for $N$ pairs of continuous variables. To find mutually unbiased bases reduces, for specific states related to the Heisenberg-Weyl group, to a problem of symplectic geometry. Given a \emph{single} pair of continuous variables, \emph{three} mutually unbiased bases are identified while \emph{five} such bases are exhibited for \emph{two} pairs of continuous variables. For $N=2$, the golden ratio occurs in the definition of these mutually unbiased bases suggesting the relevance of number theory not only in the finite-dimensional setting.
\end{abstract}


\maketitle

Mutually unbiased (MU) bases of Hilbert spaces with finite dimension $d$ (as defined by Eq. (\ref{defMUbasesford}) below) are a useful tool. If you want to experimentally determine the state of a quantum system, given only a limited supply of identical copies, the optimal strategy is to perform measurements with respect to MU bases \cite{wootters+89}. To pass a secret message to a second party, you could use quantum cryptography to establish a shared key, a procedure which relies on MU bases in the space $\mathbb{C}^2$ \cite{bennett+84,zbinden+97} or $\mathbb{C}^d$ \cite{cerf+02}. Sending a physical system carrying a spin through a noisy environment, the effect of the interactions on the state of the spin might be modelled by a specific quantum channel, conveniently described in terms of MU bases \cite{nathanson+07}. Finally, if you happen to be captured by a mean king, you might be able to meet his challenge by knowing about entangled states and MU bases \cite{aharonov+01,kimura+06}.

Many of the ideas which underlie physical concepts defined for \emph{discrete} variables, that is, in a Hilbert space of finite dimension, survive the transition from spin operators to position and momentum operators. Quantum key distribution \cite{grosshans+03} and quantum teleportation \cite{furusawa+98}, for example, possess counterparts for \emph{continuous} variables \cite{braunstein+05} which act on an infinite-dimensional Hilbert space. It is thus natural to inquire into MU bases for continuous variables which, in fact, naturally occur in Feynman's path integral formulation of quantum mechanics \cite{svetlichny07}. The properties of MU bases in an infinite-dimensional space might also provide new insights into the existence of \emph{complete} sets of MU bases in spaces of finite dimension not equal to the power of a prime.

Let us recall the definition of MU bases in $\mathbb{C}^d$ and some of their properties. Two orthonormal bases ${\cal B}_b= \{ \ket{\psi^{b}_j} \}_{j=1 \ldots d}$ and ${\cal B}_{b^\prime}=\{ \ket{\psi^{b^\prime}_j} \}_{j=1 \ldots d}$ are called MU if
\be
| \bra{\psi^{b}_j}\psi^{b^\prime}_{j^\prime} \rangle |
   = \left\{ \begin{array}{ccr}
           \delta_{jj^\prime} & \mbox{ if } & b=b^\prime \, ,\\
               \kappa >0              & \mbox{ if } & b\neq b^\prime \, ,\\
           \end{array}\right.
\label{defMUbasesford}
\ee
since each state of one basis gives rise to the same probability distribution when measured with respect to the other basis. The value of the overlap $\kappa$ is \emph{not} arbitrary but one \emph{derives} from (\ref{defMUbasesford}) that $\kappa \equiv 1/\sqrt{d}$ by using the completeness of the basis ${\cal B}_b$, say.

Schwinger \cite{schwinger60} describes how to construct two MU bases from any orthonormal basis of $\mathbb{C}^d$. They are found to be the eigenbases of two operators $\hat U$ and $\hat V$ each shifting cyclically the elements of the other basis. These operators satisfy commutation relations of Heisenberg-Weyl type, ${\hat U} {\hat V} = e^{2\pi i/d} {\hat V} {\hat U}$, describing finite translations in a discrete phase space \cite{vourdas04}. This approach has been generalized in \cite{bandyopadhyay+2001}, where it is shown that if one finds $n$ unitaries each cyclically shifting the eigenbases of all other unitaries then these $n$ bases are MU.

The number of MU bases in $\mathbb{C}^d$ is limited to $d+1$. Such \emph{complete} sets of MU bases were constructed first in the case of $d$ being a prime number \cite{ivanovic81} and subsequently for $d$ being a power of a prime \cite{wootters+89}. For composite dimensions $d=d_1 d_2\ldots d_k$, the factors being (powers of) different primes, it is currently unknown whether complete sets of MU bases exist \cite{wernerpage}. Interestingly, composite dimensions are rare for small values of $d$ but predominate for large $d$. While it is possible to construct three MU bases for any $d \geq 2$, numerical evidence for d=6 (the smallest composite integer) suggests that no four MU bases exist \cite{butterley+07} and that many of their subsets are missing as well \cite{brierley+08}.

Let us now turn to continuous variables ${\hat p}$ and ${\hat q}$, with $[{\hat q},{\hat p}]=i\hbar$, acting on the Hilbert space ${\cal L}_2(\mathbb R)$ of square-integrable functions on the real line. The (generalized) eigenstates of position and momentum $\ket{q}, q \in \mathbb{R}$, and $\ket{p}, p \in \mathbb{R}$, respectively, are known to satisfy
\be
\bra{q}p\rangle = \frac{1}{\sqrt{2\pi \hbar}}e^{iqp/\hbar} \, .
\label{pqareMU}
\ee
Thus, a natural generalization of (\ref{defMUbasesford}) for bases $\{ \ket{\psi_s^{b}} \}_{s \in \mathbb{R}}$ of an infinite-dimensional Hilbert space takes the form
\be
| \bra{\psi^{b}_s}\psi^{b^\prime}_{s^\prime} \rangle |
   = \left\{ \begin{array}{ccr}
           \delta(s-s^\prime) & \mbox{ if } & b=b^\prime \, ,\\
               k >0              & \mbox{ if } & b\neq b^\prime \, ,\\
           \end{array}\right.
\label{defMUbasesforCV}
\ee
where the $\delta$-normalization of the states reflects the fact that the labels $s,s^\prime$ are \emph{continous}. Consequently, the eigenstates of the position and momentum operators provide an example of MU bases with $k=1/\sqrt{2\pi\hbar}$. The appearance of generalized eigenstates is inevitable, because no normalizable state exists which has a non-zero overlap with all elements of a countable orthonormal basis.

Is it possible to find three or more MU bases for one pair of continuous variables? The momentum basis ${\cal B}_p$ results from a rotation of the position basis ${\cal B}_q$ by an angle $\pi/2$. Thus, a third MU basis might be given by ${\cal B}_\vartheta = \{ \ket{q_\vartheta} \}_{q_\vartheta \in \mathbb{R}}$, the eigenbasis of the operator ${\hat q}_\vartheta = {\hat q}\cos \vartheta + {\hat p} \sin \vartheta$ with eigenvalue $q_\vartheta$, $\vartheta \in (0,2\pi)$. Using Wigner functions, one finds that the modulus of the overlap between states of ${\cal B}_q$ and ${\cal B}_\vartheta$ is
\be
|\bk{q_\vartheta}{q}|^2
     = \frac{1}{2\pi \hbar \, | \! \sin \vartheta|} \neq \frac{1}{2\pi \hbar}\, .
\label{notathirdMUB}
\ee
Thus, no basis ${\cal B}_\vartheta$ with $\vartheta \in (0,\pi/2)$ combines with ${\cal B}_q$ and ${\cal B}_p$ to give a triple of MU bases.

There is, however, a \emph{symmetric} choice of operators which does provide \emph{three} MU bases. Consider the bases ${\cal B}_\pm = \{ \ket{ q_\pm}\}_{q_\pm \in \mathbb{R}}$ where ${\hat q}_\pm = {\hat q}\cos (2\pi/3) \pm {\hat p} \sin (2\pi/3)$, obtained from rotating the position basis by the angles $\pm 2\pi/3$, respectively. One finds
\be
|\bk{q}{q_+}|^2 = |\bk{q_+}{q_-}|^2 = |\bk{q_-}{q}|^2 = \frac{1}{2 \pi \hbar \,  | \! \sin(2\pi/3)|}\, ,
\label{symmetricconfiguration}
\ee
so that the triple ${\cal B}_+$, ${\cal B}_-$, and ${\cal B}_q$ \emph{is} MU with overlap $k=1/\sqrt{\pi \hbar\sqrt{3}}$ in (\ref{defMUbasesforCV}). Comparing this result with (\ref{pqareMU}), we realize that, for continuous variables, the constant $k$ in (\ref{defMUbasesforCV}) may take different values for different MU bases.

In spite of (\ref{notathirdMUB}), it is possible to complement ${\cal B}_q$ and ${\cal B}_p$ with a third basis resulting in an \emph{asymmetric} triple of MU bases. Consider ${\cal B}_{q-p}$ consisting of the eigenstates of the operator ${\hat q} - {\hat p} \equiv \sqrt{2}{\hat q}_{\pi/4}$ which \emph{cannot} be obtained from $\hat q$ by a rotation due to the factor $\sqrt{2}$. Nevertheless, one finds (as stated in \cite{delatorre+2001}) that
\be
|\bk{q}{q-p}|^2 = |\bk{q-p}{p}|^2 = |\bk{p}{q}|^2 = \frac{1}{2 \pi \hbar} \, ,
\label{asymmetric}
\ee
providing us with an \emph{asymmetric} triple of MU bases.

We now develop a systematic approach to MU bases for $N$ pairs of continuous variables residing in product states. For $N=1$, we will be able to explain the observations above. For $N\geq2$, we will derive geometric conditions which express whether product-state bases are MU or not. A set of \emph{five} MU bases will be found explicitly for \emph{two} continuous variables. Subsequently, we will formulate conditions to be MU for bases which do not have to consist of product states only.

The Heisenberg-Weyl operator
\be
\label{Weyl-Heisenberg}
\hat T({\bf a}) = \exp[i(P{\hat q} - Q{\hat p})/\hbar]
\ee
which translates  the position of a wavefunction by $Q$ and boosts its momentum by $P$, will play a central role. We consider the generator $\hat {x}_{\bf a}$ of an infinitesimal translation in the direction
${\bf a}^t=(Q,P)$, using the notation:
\be
{\hat x}_{{\bf a}} \equiv P {\hat q} - Q {\hat p}\equiv {\bf a}^t\cdot {\bf j}\cdot \hat {\bf x}
         \quad
         \mbox{with } \,
            {\bf j} = \left( \begin{array}{cc}
                            0 & -1 \\
                            1 & 0
                            \end{array} \right)
\label{definez}
\ee
where $\hat{\bf x}=(\hat q,\hat p)^t$.  Denote the eigenstates of ${\hat x}_{{\bf a}}$ by $\ket{{\bf a},\alpha}$ where ${\bf a}$ identifies a particular family of states and $\alpha$ labels an element of this family. They satisfy
\be
{\hat x}_{{\bf a}} \ket{{\bf a} , \alpha} = \alpha \ket{{\bf a} , \alpha} \, , \, \alpha \in \mathbb{R} \, ,
\label{zeigenequation}
\ee
forming complete and $\delta$-orthonormal families of states ${\cal B}_{{\bf a}}$. Their position representations are given by
\be
\bk{q}{{\bf a} , \alpha} = \frac{1}{\sqrt{2\pi\hbar|Q|}} e^{iP (q - \alpha/P)^2/2\hbar Q} \, ,
\label{posrep}
\ee
if both $P$ and $Q$ are non-zero \cite{gibbons+04}. The scalar product between states from bases with labels ${\bf a}$ and ${\bf b}$ $(\neq {\bf a})$ is found to be
\be
|\bk{{\bf b} , \beta}{{\bf a} , \alpha}|^2
         = \frac{1}{ 2\pi\hbar |{\bf b}^t \cdot {\bf j} \cdot {\bf a}|}\, .
\label{symplecticmatrix}
\ee
It is crucial for the following that the right-hand-side of (\ref{symplecticmatrix}) depends only on the modulus of the \emph{symplectic} product of the vectors ${\bf a}$ and ${\bf b}$, which is equal to the (unsigned) area of the parallelogram defined by these vectors. The particular class of states considered here thus picks up the symplectic structure related to the commutation relations $[{\hat x}_{{\bf a}},{\hat x}_{{\bf b}}]= - i \hbar {\bf a}^t\cdot{\bf j} \cdot {\bf b}$. Note that Eq. (\ref{posrep}) is consistent with (\ref{symplecticmatrix}) since one has $\ket{q}\equiv \ket{{\bf e}_P, \alpha}$, with ${\bf e}^t_P \equiv (0,1)$.

We are now in a position to derive sufficient conditions to have MU bases for $N$ pairs of continuous variables ${\hat {\bf x}}_n =({\hat q}_n, {\hat p}_n)$, $n=1 \ldots N$, with $[{\hat q}_n,{\hat p}_{n^\prime}] = i\hbar\delta_{nn^\prime}$, each pair ${\hat {\bf x}}_n$ acting on a copy of ${\cal L}_2(\mathbb{R})$.

In a first step, we restrict the candidates for MU bases to $N$-fold tensor products of the states in (\ref{zeigenequation}),
\be
\ket{\vec{\bf a} , \vec{\alpha}} \equiv \ket{{\bf a}_1 , \alpha_1}
      \otimes \ldots \otimes \ket{{\bf a}_N , \alpha_N}
      \equiv \bigotimes_{n=1}^N \ket{{\bf a}_n , \alpha_n}\, ,
\label{Nline}
\ee
which define a complete and $\delta$-orthonormal basis ${\cal B}_{\vec{\bf a}}$. Using (\ref{symplecticmatrix}), the modulus of the scalar product of $\ket{\vec{\bf a} , \vec{\alpha}}$ and $\ket{\vec{\bf b} , \vec{\beta}}$ is given by
\bea
|\bk{\vec{\bf a} , \vec{\alpha}}{\vec{\bf b} , \vec{\beta}}|^2
 &=& \prod_{n=1}^N |\bk{{\bf a}_n , \alpha_n}{{\bf b}_n , \beta_n}|^2
 \nonumber\\
 &=& \frac{1}{(2\pi\hbar)^N} \prod_{n=1}^N \frac{1}{|{\bf a}^t_n \cdot {\bf j} \cdot {\bf b}_n|} \, ,
\label{Noverlap}
\eea
which can be written as
\be
|\bk{\vec{\bf a} , \vec{\alpha}}{\vec{\bf b} , \vec{\beta}}|^2
  = (2\pi\hbar)^{-N}|\vec{\bf a}^t \cdot {\bf j}_N \cdot \vec{\bf b}|^{-1}
\label{Ntensoroverlap}
\ee
where
\be
\label{Outerprodvector}
\vec{\bf a} = {\bf a}_1 \otimes \ldots \otimes {\bf a}_N\ ,\ \ \  {\bf j}_N =
{\bf j}^{\otimes N}\, ,
\ee
etc. Thus, in order that some bases ${\cal B}_{\vec{a}}$, ${\cal B}_{\vec{b}}, \ldots$, be MU, the unsigned symplectic products between any pairs of the vectors $\vec{\bf a}, \vec{\bf b}, \ldots$ must take one and the same value,
\be
  |\vec{\bf a}^t \cdot {\bf j}_N \cdot \vec{\bf b}|
  = |\vec{\bf b}^t \cdot {\bf j}_N \cdot \vec{\bf c}|
  = \ldots = K >0 \, ,
\label{Nsymplgeometricalcondition}
\ee
reducing the search for MU bases of product form (\ref{symplecticmatrix}) to the search of product vectors $\vec{\bf a},\vec{\bf b},\ldots$ in $\mathbb{R}^{2^N}$ satisfying (\ref{Nsymplgeometricalcondition}). Having found a solution $\{ \vec{\bf a}, \vec{\bf b}, \ldots \}$ for some value of the constant $K$, one finds a solution for any other positive $K^\prime$ by rescaling each vector with the factor $\sqrt{K/K^\prime}$.

What is the maximal number of vectors satisfying (\ref{Nsymplgeometricalcondition}) for $N$ pairs of continuous variables? Lacking a general solution, we consider this problem of symplectic geometry in some detail for $N=1$ and $N=2$.

$N=1$: The constraints (\ref{Nsymplgeometricalcondition}) now read $|{\bf a}^t \cdot {\bf j} \cdot {\bf b}|
  = |{\bf b}^t \cdot {\bf j} \cdot {\bf c}|
  = |{\bf c}^t \cdot {\bf j} \cdot {\bf a}|
  = k > 0$.
In fact, only three vectors need to be written here since one can show that it is \emph{impossible} to have a fourth vector ${\bf d}$ of symplectic product $k$ with ${\bf a}, {\bf b}$ and ${\bf c}$ satisfying these conditions. This does not exclude, however, the existence of four or more MU bases built from an entirely different set of states.

Working out the unsigned symplectic product of the vectors $(0,-1)$, $(1,0)$, and $(1,1)$ leads to $k=1$, correctly reproducing the asymmetric solution presented in (\ref{asymmetric}). Similarly, the set of unit vectors $(0,-1)$ and  $(\pm\sqrt{3}/2,1)$, which is invariant under three-fold rotations, describes the \emph{symmetric} configuration (\ref{symmetricconfiguration}), with $k=\sqrt{3}/2$. These apparently different solutions are, in fact, closely related. Consider all real $2\times2$ matrices $\bf m$ with unit determinant which, under conjugation, leave the matrix ${\bf j}$ invariant up to a sign,
\be
{\bf m}^t \cdot {\bf j} \cdot {\bf m} = \pm {\bf j}\, ;
\label{unsignedsymplecticmatrix}
\ee
we will call these matrices \emph{unsigned symplectic}. They clearly form a group which consists of the union of all real symplectic $2\times2$ matrices, denoted by $\mbox{Sp}(1,\mathbb{R})$, and all these matrices multiplied by the matrix ${\bf j}$ in (\ref{definez}) which (is not symplectic but) satisfies (\ref{unsignedsymplecticmatrix}) with the minus sign. Due to (\ref{unsignedsymplecticmatrix}), symplectic products ${\bf a}^t \cdot {\bf j} \cdot {\bf b}$ remain invariant up to a sign under transformations of the form ${\bf a} \to {\bf m}\cdot {\bf a}$. Using unsigned symplectic transformations, it becomes possible to map the triple of vectors $(0,-1)$, $(1,0)$, and $(1,1)$ into a configuration with three-fold rotational symmetry which is equivalent to the three MU bases in (\ref{symmetricconfiguration}), up to a non-unitary scaling transformation as described after Eq. (\ref{Nsymplgeometricalcondition}).

$N=2$: MU bases correspond to sets of product vectors $\vec{\bf a} = {\bf a}_1\otimes {\bf a}_2$, $\vec{\bf b} = {\bf b}_1\otimes {\bf b}_2$, \ldots , with equal unsigned symplectic products. We now exhibit \emph{five} vectors which satisfy (\ref{Nsymplgeometricalcondition}) with $K=1$,
namely
\bea
                   \left( { \begin{array}{c}
                                1\\
                                0
                          \end{array}}\right) \otimes
                         \left({  \begin{array}{c}
                                1\\
                                0
                          \end{array}}\right) \, , \quad
                 \left({\begin{array}{c}
                                0\\
                                1
                          \end{array}}\right) \otimes
                          \left({\begin{array}{c}
                                0\\
                                1
                          \end{array}}\right) \, , \quad
                     \left({\begin{array}{c}
                                1\\
                                1
                          \end{array}}\right) \!\otimes
                          \!\left({\begin{array}{c}
                                1\\
                                1
                          \end{array}}\right) \, ,&& \nonumber\\
             \qquad  \left( {\begin{array}{c}
                                1\\
                                1-R
                          \end{array}}\right) \! \otimes \!
                         \left({\begin{array}{c}
                                1\\
                                R
                          \end{array}}\right) \, , \quad
                  \left({\begin{array}{c}
                               1\\
                                2-R
                   \end{array}}\right) \! \otimes \!
                          \left({\begin{array}{c}
                                1\\
                               1+R
                          \end{array}}\right) \, .\ &&
\label{fivefortwoCVs}
\eea
Here the number $R$ is the \emph{golden ratio}, i.e. the positive solution of $R^2=R+1$. Each coefficient of the five vectors is a sum of integer multiples of the numbers $1$ and $R$. Hence, we find that the coefficients are elements of a number field given by a quadratic extension of the integers (just as the field of complex numbers is an extension of the real numbers where $i$, the solution of $r^2+1=0$, plays the same role as $R$). Thus the
link between MU bases and number theory which pervades the finite-dimensional case (surveyed in, e.g.  \cite{planat+06}) also exists for continous-variables. Interestingly, MU bases for multiple qubits \cite{romero+05} or qutrits \cite{lawrence+02, lawrence04} must contain entangled states, contrary to what we find here.

In a second step, we construct MU bases for $N$ continuous variables from states not limited to the tensor products (\ref{Nline}). To do so we introduce \emph{metaplectic} operators which represent linear canonical phase space transformations in Hilbert space. Explicitly, consider the transformation ${\bf A}'={\bf M}\cdot{\bf A}$, with ${\bf A}= (q_1,\cdots,p_N) \equiv ({\bf q},{\bf p}) \in \mathbb{R}^{2N}$ and ${\bf M}$ being a symplectic matrix of size $2N \times 2N$. Then there is a unitary operator $\hat U_{\bf M}$ such that the translation operators ${\hat T}({\bf A}$)--each a product of $N$ operators of the form (\ref{Weyl-Heisenberg})--transform according to
\be
\hat U_{\bf M} \hat T({\bf A})= \hat T({\bf M}\cdot {\bf A})\hat U_{\bf M}\,,
\label{metapl defn}
\ee
defining the metaplectic $\hat U_{\bf M}$. If symplectic transformations are composed, ${\bf M}={\bf M}'\cdot {\bf M}''$, then the corresponding metaplectic operators are composed in the same manner:
$\hat U_{\bf M}=\hat U_{{\bf M}'}\hat U_{{\bf M}''}$.

The use of metaplectic operators has been implicit in our earlier discussion where we obtained a set of states $\ket{{\bf a},\alpha}$, satisfying (\ref{zeigenequation}), which are MU with respect to the position eigenstates $\ket{q}$. We now show that these states can be obtained directly by application of a metaplectic operator. Expand (\ref{metapl defn}) in ${\bf A}$ and consider the linear term to obtain
$\hat U_{\bf M}\hat x_{\bf A}=\hat x_{{\bf M}\cdot {\bf A}}\hat U_{\bf M}$.
First, let $N=1$ and choose the symplectic matrix ${\bf m}$ such that ${\bf m}\cdot {\bf a}=(0,1)^t$, so
$\hat x_{{\bf m}\cdot{\bf a}}=\hat q$.  The eigenfunctions of ${\hat x}_{\bf a}$ in (\ref{zeigenequation}) are then generated by  $\ket{{\bf a},\alpha}=\hat U_{\bf m}\ket{q}$. The symplectic matrix satisfying ${\bf m}\cdot (Q,P)^t=(0,1)^t$ is
\be
\label{SpecialM}
{\bf m}=\left(\begin{array}{cc}
         1&0\\
         \mu&1\\
         \end{array}\right)
         \left(\begin{array}{cc}
         P&-Q\\
         1/Q&0\\
         \end{array}\right)
\ee
where $\mu \in \mathbb{R}$ parametrises a shear along the line defining the states $\ket{{\bf a},\alpha}$. It affects the phase of $\bk{q}{{\bf a},\alpha}$, but not its magnitude.

In order to discuss a more general construction of MU bases (with $N\ge 1$) we use a general expression \cite{mehlig+01} for a metaplectic operators which correspond to a  symplectic matrix ${\bf M}$ of dimension $2N$,
\be
\label{Metaplectic}
\hat U_{\bf M}\!=\!\frac{\exp(i\Theta)}{\sqrt{\vert{\rm det}({\bf M}-{\bf I})\vert}}\!\!
\int \!\!\frac{d{\bf A}}{(2\pi \hbar)^N}\exp\!\left[\frac{i}{2\hbar}{\bf A}^t\cdot {\bf N}\cdot {\bf A}\right]\!
\hat T({\bf A});
\ee
here $\Theta $ is a phase which need not concern us further,
${\bf N}=\tfrac{1}{2}{\bf J}({\bf M}+{\bf I})({\bf M}-{\bf I})^{-1}$
is a symmetric matrix, ${\bf J}={\bf j}\oplus \ldots \oplus {\bf j}$ a block diagonal generalization of ${\bf j}$ in (\ref{definez}), and the integration is over the $2N$ dimensions of phase space, $d{\bf A}=dQ_1 \, dQ_2 \ldots dP_N$. The matrices ${\bf M}$ and ${\bf N}$ may be written using  blocks of dimension $N\times N$,
\be
\label{Blockmatrices}
\left(\begin{array}{c}{\bf q}'\\{\bf p}'\\
\end{array}\right)=
\left(\begin{array}{cc}{\bf M}_{qq}&{\bf M}_{qp}\\{\bf M}_{pq}&{\bf M}_{pp}\\\end{array}\right)
\left(\begin{array}{c}{\bf q}\\{\bf p}\\\end{array}\right)\ ,\ \ \
{\bf N}=\left(\begin{array}{cc}{\bf N}_{qq}&{\bf N}_{qp}\\{\bf N}_{pq}&{\bf N}_{pp}\end{array}\right) \, .
\ee
Consider the action of $\hat U_{\bf M}$ on $N$-fold products of position eigenstates, $\ket{{\bf q}} \equiv \ket{q_1} \otimes \ldots \otimes \ket{q_N}$. Using (\ref{Metaplectic}) and (\ref{Blockmatrices}), we find that states $\hat U_{\bf M}\ket{{\bf q}} \equiv \ket{{\bf M},{\bf q}}$ are unbiased relative to the position eigenstates, i.e.,
\be
\label{GenMU}
\vert \langle {\bf q}'\vert {\bf M},{\bf q} \rangle\vert^2=\frac{1}{(2\pi \hbar)^N}\frac{1}{\vert {\rm det}({\bf M}-{\bf I}){\rm det}({\bf N}_{pp})\vert}\ \, .
\ee
It follows from the composition property of metaplectic matrices 
that states $\ket{{\bf M},{\bf q}}$ with different ${\bf M}$ are, and that the magnitude of their overlap can be calculated by composing the underlying symplectic matrices:
\bea
\label{MoregenMU}
\vert\bk{{\bf M},{\bf q}}{{\bf M}',{\bf q}'}\vert^2
 &=& \vert\bk{{\bf q}}{\hat U_{\bf M}^{-1}\hat U_{{\bf M}'}
                              \vert {\bf q}'} \vert^2\nonumber \\
 &=& \vert\bk{{\bf q}}{({\bf M}^{-1}{\bf M}'), {\bf q}'}\vert^2\, ,
\eea
where the final expression is evaluated using (\ref{GenMU}). Thus, the problem of finding MU bases associated with metaplectic operators can be solved by finding symplectic transformations such that the resulting expressions on the right-hand-side of (\ref{GenMU}) take the same values.
This may allow for a much larger set of MU bases than (\ref{Nsymplgeometricalcondition}).

Our principal results are conditions for bases related by a metaplectic transformation to be MU, namely (\ref{Nsymplgeometricalcondition}) (for which we found a solution (\ref{fivefortwoCVs})) and more generally (\ref{MoregenMU}) (as yet unexplored).
To conclude we point out open questions. Even in the case of $N=1$, it is not know whether more than three MU bases exist. To have only three MU bases would be slightly surprising as the limit of $d \to \infty$ passing through prime dimensions suggests the existence of an  unlimited number of MU bases. The result (\ref{notathirdMUB}) confirms this expectation in a restricted sense--any \emph{pair} of bases ${\cal B}_\vartheta$ and ${\cal B}_{\vartheta^\prime}$ is MU but with possibly \emph{different} values for the overlap. Future studies will reveal whether the pairwise unbiased bases ${\cal B}_\vartheta, \vartheta \in(0,\pi/2)$ are as useful as a \emph{complete} set of MU bases.

It is also unknown whether the bases ${\cal B}_q$ and ${\cal B}_p$ can be supplemented by a third MU basis \emph{qualitatively different} from the one presented in (\ref{asymmetric}). Let the state $\ket{\psi}$ be a member of such a basis. The conditions $|\bk{q}{\psi}| = |\bk{p}{\psi}| = 1/\sqrt{2\pi\hbar}$ imply that its expansion coefficients in the position and momentum basis are constant multiples of phase factors $\exp [if(q)]$
 and $\exp [ig(p)]$, respectively, related to each other by a Fourier transform,
\be
e^{ig(p)} = \frac{1}{\sqrt{2\pi\hbar}} \int_{-\infty}^\infty e^{if(q)} e^{-ipq/\hbar} dq \, .
\label{integraleq}
\ee
Thus, if the only pairs of functions $(f(q),g(p))$ solving this integral equation consist of quadratic polynomials, then there are no MU bases beyond the ones exhibited so far. Unfortunately, the entire set of its solutions is not known to us.

{\bf Acknowledgements:}
We thank Tony Sudbery for his comments and the London Mathematical Society for financial support through a Scheme 4 Grant (Ref. 4625).

\end{document}